# Technical support for Life Sciences communities on a production grid infrastructure

Franck MICHEL[a], Johan MONTAGNAT[a], Tristan GLATARD[b]
[a] *CNRS/UNS, I3S laboratory, MODALIS team, 06903 Sophia Antipolis, FRANCE*
[b] *Université de Lyon, CNRS, INSERM, CREATIS, 69621 Villeurbanne, FRANCE*

**Abstract.** Production operation of large distributed computing infrastructures (DCI) still requires a lot of human intervention to reach acceptable quality of service. This may be achievable for scientific communities with solid IT support, but it remains a show-stopper for others. Some application execution environments are used to hide runtime technical issues from end users. But they mostly aim at fault-tolerance rather than incident resolution, and their operation still requires substantial manpower. A longer-term support activity is thus needed to ensure sustained quality of service for Virtual Organisations (VO). This paper describes how the *biomed* VO has addressed this challenge by setting up a technical support team. Its organisation, tooling, daily tasks, and procedures are described. Results are shown in terms of resource usage by end users, amount of reported incidents, and developed software tools. Based on our experience, we suggest ways to measure the impact of the technical support, perspectives to decrease its human cost and make it more community-specific.

## 1. Introduction, motivations

The European Grid Initiative (EGI) delivers a sustainable worldwide production grid infrastructure to scientists organised in Virtual Research Communities (VRCs). For those communities, operating large subsets of this infrastructure with reasonable quality of service requires substantial efforts to cope with runtime issues such as configuration flaws, hardware failures, load balancing, or storage unavailability. Early grid adopters such as High Energy Physics (HEP) are able to invest the required effort to solve these issues [7], but this is out of reach of communities that are more fragmented, or that simply cannot afford such a costly IT support. Therefore, operating a Virtual Organisation (VO) with reduced human cost has become a critical concern for the future of grids in academic research.

To address this challenge, some communities have developed application execution environments such as pilot-job systems [1] [4] [8], workflow managers [5], and portals [2] [3] [9] [10], that overlay core middleware to ensure fault tolerance and efficiency of week-long experiments. They provide features such as queuing time reduction, dispatching of tasks on multiple infrastructures, file replica management and resources black/white-listing. This approach is efficient to help experiments complete successfully, but it also comes with concerns. Firstly, it is difficult to enforce their adoption in fragmented communities with heterogeneous requirements. Secondly, they hide technical issues but do not solve them. In particular, a technical issue may affect only a single VO, and it may last endlessly unless someone from that VO manually points it out (e.g. quota policies, authentication issues, or misconfigurations). Consequently, a longer-term, VO-specific but application-independent support activity is needed. This support should act on behalf of a large community of users and applications, to ensure a sustained quality of service at the grid infrastructure level. The complementarity of application execution and support environments is illustrated in Figure 1. Within the Life Science Grid Community[1] (LSGC), dozens of independent users applications generate all kinds of computing tasks. There is no control on the type of grid usage, and the use of a single execution environment that could simplify VO management is not possible.

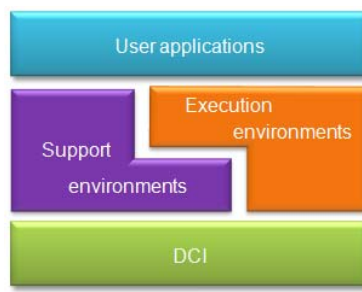

**Figure 1.** Complementarity of support and application execution environments

This paper presents the activities and achievements of the technical support team of *biomed,* an international VO of the LSGC, that addresses bioinformatics, drug discovery and medical imaging. Resource monitoring, error reporting and requirement collection are its main levers to ensure that 295 registered users can properly use resources from 150 sites worldwide. Results are shown in terms of resource usage by end users, amount of reported incidents, and monitoring software tools developed. Ways of measuring the impact of the support activity are discussed, as well as perspectives to decrease its human cost and thus allow the team to move on to more life-science specific support.

## 2. Technical context of VO operations

Various tools have been developed to monitor resources in DCIs. This section provides a categorized list of examples from the EGI ecosystem.

**Topology and information services**: topology refers to the list of resources and services available in the infrastructure. The Grid Operations Center Database [3] (GOCDB) provides a static view of the resources while the Berkeley Database Information Index[4] (BDII) is a distributed database that holds dynamic information about the status of all resources and services in the grid. Both are critical sources of information for the services described hereafter.

---

[1] http://wiki.healthgrid.org/LSVRC:Index
[3] https://goc.egi.eu/
[4] https://twiki.cern.ch/twiki/bin/view/EGEE/BDII

**Usage accounting**: CESGA[5] is the official EGI Accounting Portal. It collects CPU consumption data and number of jobs processed from Computing Elements (CE), but job success rates are not considered. It can break down usage data per individual user (top 10 users only), per country or per VO. CESGA is able to report detailed history data on any period of time. The Distributed Grid Accounting System[6] (DGAS) is an alternate initiative deployed in the Italian grid. It provides a stand-alone infrastructure for computing and storage accounting of countries and VOs. It provides its own sensors for many different Local Resource Management Systems, and repositories for persistent storage of usage records.

**Operations and user support**: operations refer to the identification and follow-up of incidents, the reporting to the appropriate sites or national grids, and the discussion of possible fixes. Commonly, operations are performed by teams relaying each other during duty shifts. Some tools are available to help teams follow-up on issues and pass the information along to the next one. The EGI Operations Portal[7] is dedicated to NGI support teams: it filters and classifies alarms from the Nagios test execution framework, provides resources availability charts, makes an easy link between alarms and GGUS[8] tickets, and provides ways to report the status from one team to another. The VO Admin Portal[9] integrates, on a single portal, views from various other portals, with a VO focus: EGI Operations Portal, CESGA, GGUS, GSTAT, GOCDB, MyEGI, VOMS.

**Resources monitoring**: the Service Availability Monitoring [10] (SAM) is an extensible monitoring framework including the Nagios test execution tool, probes dedicated to monitoring resources, and the MyEGI visualization portal. The VO SAM provides customised probes to monitor specific VO resources like the Virtual Organisation Management Server (VOMS) or the LCG File Catalogue (LFC). GSTAT[11] is a visualization portal for data published in the BDII. It breaks down information per site, and reports aggregated data on storage, CPU consumption and number of jobs. It can also focus on resources supporting a particular VO. Conversely to SAM that mostly reports data from its own probes, GSTAT reports data published by the resources themselves. As individual resources may be misconfigured or rely on different interpretations of the GLUE schema specification, GSTAT information must be interpreted with care. GSTAT does not keep track of all detailed data it collects, but it is able to report long-term history data, the older the data, the coarser the grain.

The CERN Experiment Dashboard performs both at the application and infrastructure levels. It correlates multiple monitoring sources including SAM and experiments-specific services. Its *job monitoring* service keeps track of any submitted job and stores indicators such as its status, resource usage, application robustness and data access quality. The *site reliability* service estimates the site performance regarding job processing and data transfer. The dashboard also provides experiment-specific services. For instance, the *task monitoring* aggregates information about tasks (coherent sets of jobs that make sense to the physicist), while the *ATLAS data*

---

[5] http://www4.egee.cesga.es/accounting/egee_view.php
[6] http://www.to.infn.it/dgas/
[7] http://operations-portal.egi.eu/
[8] https://ggus.eu/pages/home.php
[9] https://vodashboard.lip.pt/
[10] https://wiki.egi.eu/wiki/SAM
[11] http://gstat.egi.eu/

*management monitoring* steers the large-scale distribution of the experiment data sets. Other services perform production monitoring and accounting. The CERN dashboard thereby performs at the application level by monitoring jobs and data transfers, as well as at the infrastructure level by monitoring resources and sites. Nevertheless, as discussed in section 5, these features come with several constraints that may not be easily met by other communities.

The profusion of tools such as those described above outlines how operating large DCIs is complex. This also suggests that different communities have different operational needs. Thus, each community has to assess how relevant each tool is with regards to their specific needs. This question is discussed in section 4.3.

**3. The biomed support team**

*3.1. Organisation and tools*

The *biomed* technical support team acts at the interface between grid application users and resource providers. It has been active since March 2010, and consists of eight teams of volunteer grid experts from the most active user groups. In 2011, two user groups have joined the support team while two have requested to remain as a backup only, resulting in a yearly turnover of 25% of the teams, and 21% of the members. Participating teams relay each other during duty shifts. Each team is on duty once every twelve weeks; the daily workload is estimated to 1 to 2 hours. This allows for a sustained level of support while requiring a reasonably limited burden from contributors. The coordination of relaying teams is carried out during take-over phone conferences.

A "SAM box" (Service Availability Monitoring) is dedicated to the *biomed* VO. Its Nagios instance monitors single point of failure services like the LFC and VOMS, as well as all storage and computing resources available to the VO. Incident reports and follow-up are shared by team members using GGUS *team* tickets. Site administrators may also submit *VO Support* tickets to communicate with the support team, e.g. to announce SE decommissioning, or to report excessive CPU consumption or suspect behaviour of a user. In this case, the support team liaises with the users and takes appropriate measures.

Some non VO-specific tools are used on a daily basis by the support the team: GOCDB, BDII, VOMS, CESGA, GSTAT. Some members of the support team are surveying other tools and portals to improve support efficiency while reducing its workload. New tools are eventually developed to implement features that are not covered yet (section 4.4 provides more details).

*3.2. Tasks and procedures*

Duties of the support team focus on generic monitoring activities and, until now, do not span VO specific applications. Daily tasks, procedures and advice on how to deal with issues are documented on a dedicated support team wiki[12]. The team on duty identifies and follows up on issues, discusses salient technical problems, and investigates solutions with administrators.

---
[12] http://wiki.healthgrid.org/Biomed-Shifts:Practices

Data unavailability has critical consequences for users as it is often responsible for job failures. Therefore, the support team specifically focuses on Storage Elements (SE) monitoring to anticipate potential issues. Detailed procedures address the scheduled decommissioning of an SE, and the cleanup of SEs with little remaining free space. Those activities prove to be time consuming as they involve several manual steps and interactions with administrators and users. They are also hampered by "zombie" and "ghost" files: "zombies" are files stored on SEs but not registered in any file catalogue, whereas "ghosts" are catalogue entries with no corresponding physical replica. Tools such as DIRAC periodically run agents to check consistency between catalogues and storage, but such operations remain very heavy. Procedures and best practices are being improved based on this experience.

The *biomed* support team also gathers and conveys technical concerns from the users community to the EGI instances. Requirements for evolutions of the grid middleware are reported in the EGI Requirement Tracker[13] and discussed with the EGI User Community Support Team[14] (UCST), as well as questions related to services deployment and hosting, and the sharing and promoting of existing initiatives. The support team is also represented by the LSGC during the EGI User Community Board (UCB) conferences.

**4. Results and discussion**

*4.1. Resource usage*

In 2011, *biomed* consumed 19 millions normalized CPU hours, which makes it the most active life-science VO in EGI. Used computing resources are located in 22 countries or regions, and generally shared with other VOs. The monthly consumption mostly shows a series of a few peaks of activity, probably resulting from large experimental campaigns. Thereby, no trend can be observed for now, whereas we could expect a consumption increase along with the VO growth.

The average ratio of waiting jobs over running jobs is 3.9. Although this value should be interpreted with caution (see section 4.3), such a high ratio is an explanation for long delays in the job queues. Increasing the set of supporting computing resources, in particular dedicated ones, is required to improve the overall performance of *biomed* jobs. Besides, a survey has just started to assess CEs reliability by other ways than SAM probes. It appears that a significant number of CEs, although not actually faulty, cannot process jobs in a reasonable time. Work is ongoing to find out possible causes.

During the same period, the used storage space has raised from approximately 1.2 PB to 2 PB, out of a total 3.7 PB. A scan of all SEs supporting *biomed* revealed that users usually do not clean up their files when they leave the VO. As a result, the total space used by the VO keeps rising. To address this concern, a clean-up campaign will be initiated in 2012.

Local inspection of CPU and storage consumption data shows that the usage of resources varies significantly from one site to another. The reasons of this variability are probably manifold, however one at least can be outlined: due to the low reliability of some sites, users tend to target specifically well-known reliable resources (possibly local resources, or following the advice of other researchers). In addition, application execution environments often provide the ability to create white lists of resources, thus hampering the possibility to discover and use other resources. Consequently, while the average reliability of resources may eventually rise, the range of used resources remains rather stable.

---

[13] https://rt.egi.eu/rt/
[14] http://www.egi.eu/user-support

*4.2. Reported incidents*

The *biomed* support team currently monitors all resources supporting the VO: 108 Storage Elements (SE), 186 Computing Elements (CE) and 36 Workload Management Systems (WMS), a redundant VOMS server, and an LFC server. From March 2010 to December 2011 it has submitted 415 GGUS tickets (5.1 tickets per week in average). Figure 2 shows the distribution of tickets by month and type. Category *User* gathers VO Support tickets submitted by resources administrators, while category *Other* gathers tickets concerning services like VOMS, LFC, or any other question.

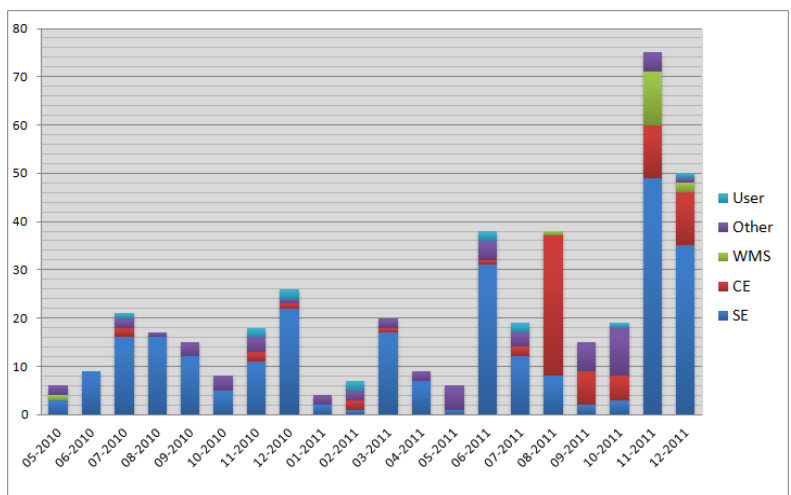

**Figure 2.** Number of GGUS tickets submitted per month and per type of issue

Different types of issues have been progressively monitored by the support team: SEs have been monitored from May 2010, CEs from August 2011, and WMSs from November 2011. Procedures to handle full SEs or erroneous data published by SEs in the BDII have been introduced in November and December 2011.

*4.3. Accuracy of support tools*

A difficulty commonly encountered by support team members is the relative reliability of monitoring data provided by support tools. In 2011, a manual review showed that 16% of the *biomed* SEs either published erroneous data in the BDII or were misconfigured, but none of those issues were detected by Nagios. Such problems result in various behaviours such as wrong estimations of storage space used by the VO, or full SEs still reporting free space. Similar issues were identified in jobs and CPU figures published by CEs. To date, approximately 10% of CEs report invalid data. Erroneous values corrupt the overall figures, making it difficult to estimate how trustful the consolidated figures computed by GSTAT are.

The EGI Accounting Portal (CESGA) is another example of how a visualization portal is sensitive to the quality of the data it relies on. Due to legal or technical constraints, some countries do not report complete user information, while some do not report jobs launched by local users. As a result, the individual and per-VO accounting data lacks some inputs, making values uncertain. Nevertheless, CESGA is considered to provide a fair CPU consumption trend, even if not totally accurate.

It is not yet possible to precisely assess the accuracy of the information provided by the MyEGI visualization portal, because it is currently not appropriately fed with data from the resources supporting the VO. Experts currently investigate the problem.

The accuracy of support tools does not only relies on the data published by resources, but also on the list of resources the VO chooses to use, i.e. the VO resources topology. MyEGI uses per-VO profiles, while the Nagios test framework relies on the BDII dynamic (and thereby versatile) state and the GOCDB static state to come up with a list of resources to monitor. None of these lists allow for any customization. The VO feeds mechanism, developed as part of the SAM framework and used by HEP VOs, specifically addresses this concern, and the support team holds an ongoing discussion on the interest of using it for *biomed*.

*4.4. Tooling improvement*

A continuous work is being performed to figure out metrics that measure resource quality of service, and to improve support procedures. With this goal, several tools have recently been developed, or are currently being developed:

(1) A tool[15] computes the filling rate of SEs every 30 minutes, and reports results along various sorting modes. The *biomed* support team uses it to detect abnormal filling rates or BDII data publication issues, while users can use it to be aware of the most and least loaded SEs.

(2) To prevent SEs from getting full, or to apply curative actions when this happens, a tool[16] regularly scans SEs which filling rate is over 80%, lists heaviest users and generates templates of email notifications to users. This relies on the *LFCBrowseSE* tool developed by the University of Valencia.

(3) The EGI Operations Portal was designed as a site-centric monitoring tool for site administrators and support teams. In 2011, the *biomed* support team and the portal developers agreed on the scope of evolutions that would be required for this portal to provide a VO-centric view, taking *biomed* as a typical use case. The resulting *VO Operations Dashboard* is currently under development and shall be released by mid-2012. It is expected to become a daily operation tool for the *biomed* support team.

Experience collected so far has helped delineate features that would improve support quality and efficiency, but that no tool covers so far. Some features have been addressed by the LSGC, while some are being discussed with UCST or are provided by other communities. Leveraging the experience of the *biomed* support team as well as other life science VOs, some more specific features have been gathered in the specification of the so-called *LSGC Dashboard*, still to be developed. It is expected to cover the following features:
- Handling of users VO-membership and group-membership registration life-cycle ;

---

[15] http://wiki.healthgrid.org/Biomed-Shifts:Index#Status_of_Biomed_Online_Storage_Space
[16] https://grid.unice.fr/~fmichel/monitor-se/reports-history.php

- Support for robot certificates in order to identify actual users 'behind' the certificate;
- Maintain mailing lists for VO users and thematic VO sub-groups;
- Collect feedback on the infrastructure, scientific production (publications);
- Automate files migration and SE decommissioning procedures, automate clean up of files whose owners have signed off the VO;
- Compute specific accounting metrics: global LSGC resource usage, per VO resource usage, per VO sub-group resource usage;
- Better advertising of scheduled downtimes to VO users;
- Maintain and advertise the topology of resources supporting the VO.

These developments should help the support team to improve its efficiency. Automating and mutualising generic operational tasks among several VOs would save time spent on generic activities, and allow the team to focus on more community-specific activities: improve front-line support, encourage the adoption of common frameworks and practices, provide application-specific support.

*4.5. Measuring the impact of the support activity*

After one year and a half operating, it is worth questioning the impact of the support team on the reliability of the VO, as perceived by *biomed* users. But coming up with appropriate metrics to quantify the return on investment remains difficult. Various criteria can potentially reflect the team's activity, thus helping to come up with significant metrics: (i) evolution of resource consumption and reliability, (ii) pace of submitted and received GGUS tickets, (iii) evolution of user satisfaction. These criteria are discussed below.

(i) Section 4.1 has shown that no evolution of the CPU consumption can be evidenced. This results from the short hindsight we have so far, as compared to the time scale of experimental campaigns. As to the used storage space, reasons for its growth can be manifold, and cannot be interpreted as an indicator of some possible improvement of SEs reliability. Besides, measuring changes in the reliability of the resources requires history reports of the number of alarms, and up and down times. All EGI sites publish availability and reliability reports, but to our knowledge, there is no way so far to produce reliability reports spanning only the resources that support a VO. Therefore, it remains difficult to figure out, from a VO perspective, how the resources reliability evolves.

(ii) A first approximation could assume that the more efficient the support activity, the more reliable the resources, and in turn the less GGUS tickets submitted. Although results presented in section 4.2 definitely reflect the important work achieved, again the hindsight is not sufficient so far to come up with a trend that would eventually reflect resources reliability improvement. Until now, the observed increasing number of GGUS tickets mostly attests of the ramping up of the support team activity.

(iii) User satisfaction is another interesting metrics, although it is difficult to measure objectively. Furthermore, it may result from changes independent of the support team's activity: for instance, changing the policy for selecting resources in platforms like DIRAC, VIP or OpenMOLE may result in a higher rate of successful jobs thereby increasing the satisfaction of users, but not being related to the support team's activity in any way.

Finally, the criteria considered above should be of interest when the hindsight will be sufficient to come up with significant trends. For the time being though, they do not prove to be meaningful. Besides, it must be noted that the support team is part of a very large ecosystem, where many actors can influence the way things work; in this context, making absolutely sure that a cause has actually produced a given measured effect is probably illusory.

**5. Conclusions**

The *biomed* technical support team has been operating for almost two years, receiving the sustained contribution from 8 institutions worldwide. Based on monitoring tools provided by EGI, its core task is to report operational incidents related to storage elements, computing elements, workload management services, file catalogues and VO management services. It applies specific procedures in case of decommissioned or full storage. It liaises as a technical interface between VO users and resource providers when user-specific issues occur. In average, 5 incidents are reported every week; each incident is solved within 14 days, includes 10 steps and involves 3.5 people. The quality of service achieved in *biomed* allows 295 registered users to yearly consume 19 millions CPU hours from 150 sites, making it the most active life-science VO on EGI in terms of CPU time.

This has limitations though. First, the daily workload required from support team members on duty is high, and the volunteering model has to be questioned as the community grows. Typically, being on duty once every 3 months is a reasonable load, but it is not sufficient for members to get fully used to procedures, and to be aware of current issues. Each shift takeover requires a costly "re-learn curve", and ultimately, the coordination of a collection of independent international teams proves to be challenging. Second, the scope of the support activities focuses on very generic operational tasks, whereas the community could benefit from VO-specific activities like the support of usual life sciences applications. To tackle this concern, it should be possible to mutualise efforts on generic tasks with other VOs of the community, e.g. by gathering monitoring activities in a single multi-VO SAM box.

Below we discuss several options to reduce the overall support cost in the middle term, and improve its efficiency. Better integrating application execution environments and support tools is a considered future direction. Among other tools, the CERN Experiment Dashboard stands at the edge of both. Reusing it for life sciences is not easy as it comes with constraints that the community may not be ready to adopt: (i) job submission tools must be instrumented to report data to the CERN dashboard, unless a supported pilot job submission system is used; (ii) site administrators use it to monitor their own site, which is not common in life sciences; (iii) while the core of the dashboard is rather generic, it remains to be studies how HEP specific developments could be ported to other communities. Nevertheless, the convergence of heterogeneous applications on a few common frameworks would certainly help users to improve the efficiency of their applications: for instance, a VO-dedicated instance of a generic framework like DIRAC could be deployed and maintained by the VO support team. This approach would improve the support efficiency by mutualising efforts. Simultaneously it would help fulfil VO accounting needs that are not addressed today.

Another initiative, like the SAM VO feeds, could help VO managers to select resources that currently support the VO on a customised basis. Matched against other data sources like downtimes or filling rate of SEs, it could provide a very accurate and up to date white list of "VO-certified reliable" resources. But to reach its full potential, this mechanism should span from low-level support tools to application execution environments.

Ultimately, it appears that sticking to the total independence of users in terms of applications design choices may not be sustainable. In the long term, the life-science community may encourage the use of common frameworks and practices, and enforce common policies. A distributed support team could then ensure a sustained high level of reliability at a lower cost.

## 6. Acknowledgements

The *biomed* technical support team is based on volunteering from most active user groups. We warmly thank its members from the following institutions and laboratories: BME-IIT (Hungary), CNRS Creatis (France), CNRS I3S (France), CNRS IPHC (France), CNRS LPC (France), CNRS ISC-PIF (France), IFI (Vietnam), INFN BA (Italy), Universitat Politècnica de València (Spain). We thank the GRIF (Grille Ile de France) for operating the *biomed* SAM box. We are grateful to EGI.eu for providing the grid infrastructure, to the support teams dealing with GGUS tickets and providing precious technical inputs, and to our active contacts at the User Community Board and the User Community Support Team. We also thank our colleagues at France-Grilles for their continuous support and useful inputs.